
  \input miniltx
  \def\Gin@driver{pdftex.def}
  \input color.sty
  \input graphicx.sty
  \resetatcatcode

%
%

%
%
%
%

\def\Serif{cmr}
\def\SerifBold{cmbx}
\def\SerifItalics{cmti}
\def\SerifSlanted{cmsl}
\def\SerifBoldItalics{cmbxti}
\def\SansSerif{cmss}
\def\SansSerifBold{cmssbx}
\def\SansSerifItalics{cmssi}
\def\SansSerifSlanted{cmssi}
\def\Math{cmmi}
\def\Symbols{cmsy}
\def\MathBold{cmmib}
\def\MoreSymbols{cmex}
\def\Typewriter{cmtt}
\def\Gothic{eufm}
\def\Double{msbm}
\def\Relazioni{msam}

= 			\Serif10 			at 5pt
= 		\SerifBold10 		at 5pt
= 	\SerifItalics10 	at 5pt
=		\SerifSlanted10 	at 5pt
=	\SerifBoldItalics10	at 5pt
= 		\SansSerif10 		at 5pt
=	\SansSerifBold10	at 5pt
=	\SansSerifItalics10	at 5pt
=	\SansSerifSlanted10	at 5pt
=				\Math10				at 5pt
=			\MathBold10			at 5pt
=			\Symbols10			at 5pt
=		\MoreSymbols10		at 5pt
=		\Typewriter10		at 5pt
=			\Gothic10			at 5pt
=			\Double10			at 5pt

= 			\Serif10 			at 7pt
= 		\SerifBold10 		at 7pt
= 	\SerifItalics10 	at 7pt
=	\SerifSlanted10 	at 7pt
=\SerifBoldItalics10	at 7pt
= 		\SansSerif10 		at 7pt
= 	\SansSerifBold10 	at 7pt
=\SansSerifItalics10	at 7pt
=\SansSerifSlanted10	at 7pt
=			\Math10				at 7pt
=		\MathBold10			at 7pt
=			\Symbols10			at 7pt
=		\MoreSymbols10		at 7pt
=		\Typewriter10		at 7pt
=			\Gothic10			at 7pt
=			\Double10			at 7pt

= 			\Serif10 			at 8pt
= 		\SerifBold10 		at 8pt
= 	\SerifItalics10 	at 8pt
=	\SerifSlanted10 	at 8pt
=\SerifBoldItalics10	at 8pt
= 		\SansSerif10 		at 8pt
= 	\SansSerifBold10 	at 8pt
=\SansSerifItalics10 at 8pt
=\SansSerifSlanted10 at 8pt
=			\Math10				at 8pt
=		\MathBold10			at 8pt
=			\Symbols10			at 8pt
=		\MoreSymbols10		at 8pt
=		\Typewriter10		at 8pt
=			\Gothic10			at 8pt
=			\Double10			at 8pt

= 			\Serif10 			at 10pt
= 		\SerifBold10 		at 10pt
= 		\SerifItalics10 	at 10pt
=		\SerifSlanted10 	at 10pt
=	\SerifBoldItalics10	at 10pt
= 		\SansSerif10 		at 10pt
= 	\SansSerifBold10 	at 10pt
= 	\SansSerifItalics10 at 10pt
= 	\SansSerifSlanted10 at 10pt
=				\Math10				at 10pt
=			\MathBold10			at 10pt
=			\Symbols10			at 10pt
=		\MoreSymbols10		at 10pt
=		\Typewriter10		at 10pt
=			\Gothic10			at 10pt
=			\Double10			at 10pt
=			\Relazioni10			at 10pt

= 				\Serif10 			at 12pt
= 			\SerifBold10 		at 12pt
= 		\SerifItalics10 	at 12pt
=		\SerifSlanted10 	at 12pt
=	\SerifBoldItalics10	at 12pt
= 			\SansSerif10 		at 12pt
= 		\SansSerifBold10 	at 12pt
= 	\SansSerifItalics10 at 12pt
= 	\SansSerifSlanted10 at 12pt
=				\Math10				at 12pt
=			\MathBold10			at 12pt
=			\Symbols10			at 12pt
=		\MoreSymbols10		at 12pt
=			\Typewriter10		at 12pt
=				\Gothic10			at 12pt
=				\Double10			at 12pt

= 			\Serif10 			at 14pt
= 		\SerifBold10 		at 14pt
= 	\SerifItalics10 	at 14pt
=		\SerifSlanted10 	at 14pt
=	\SerifBoldItalics10	at 14pt
= 		\SansSerif10 		at 14pt
= 	\SansSerifBold10 	at 14pt
= \SansSerifSlanted10 at 14pt
= \SansSerifItalics10 at 14pt
=				\Math10				at 14pt
=			\MathBold10			at 14pt
=			\Symbols10			at 14pt
=		\MoreSymbols10		at 14pt
=		\Typewriter10		at 14pt
=			\Gothic10			at 14pt
=			\Double10			at 14pt

\def\NormalStyle{\parindent=5pt\parskip=3pt\normalbaselineskip=14pt%
\def\nt{\tenSerif}%
\def\rm{\fam0\tenSerif}%
\textfont0=\tenSerif\scriptfont0=\sevenSerif\scriptscriptfont0=\fiveSerif
\textfont1=\tenMath\scriptfont1=\sevenMath\scriptscriptfont1=\fiveMath
\textfont2=\tenSymbols\scriptfont2=\sevenSymbols\scriptscriptfont2=\fiveSymbols
\textfont3=\tenMoreSymbols\scriptfont3=\sevenMoreSymbols\scriptscriptfont3=\fiveMoreSymbols
\textfont\itfam=\tenSerifItalics\def\it{\fam\itfam\tenSerifItalics}%
\textfont\slfam=\tenSerifSlanted\def\sl{\fam\slfam\tenSerifSlanted}%
\textfont\ttfam=\tenTypewriter\def\tt{\fam\ttfam\tenTypewriter}%
\textfont\bffam=\tenSerifBold%
\def\bf{\fam\bffam\tenSerifBold}\scriptfont\bffam=\sevenSerifBold\scriptscriptfont\bffam=\fiveSerifBold%
\def\cal{\tenSymbols}%
\def\greekbold{\tenMathBold}%
\def\gothic{\tenGothic}%
\def\Bbb{\tenDouble}%
\def\LieFont{\tenSerifItalics}%
\nt\normalbaselines\baselineskip=14pt%
}

\def\TitleStyle{\parindent=0pt\parskip=0pt\normalbaselineskip=15pt%
\def\nt{\fourteenSansSerifBold}%
\def\rm{\fam0\fourteenSansSerifBold}%
\textfont0=\fourteenSansSerifBold\scriptfont0=\tenSansSerifBold\scriptscriptfont0=\eightSansSerifBold
\textfont1=\fourteenMath\scriptfont1=\tenMath\scriptscriptfont1=\eightMath
\textfont2=\fourteenSymbols\scriptfont2=\tenSymbols\scriptscriptfont2=\eightSymbols
\textfont3=\fourteenMoreSymbols\scriptfont3=\tenMoreSymbols\scriptscriptfont3=\eightMoreSymbols
\textfont\itfam=\fourteenSansSerifItalics\def\it{\fam\itfam\fourteenSansSerifItalics}%
\textfont\slfam=\fourteenSansSerifSlanted\def\sl{\fam\slfam\fourteenSerifSansSlanted}%
\textfont\ttfam=\fourteenTypewriter\def\tt{\fam\ttfam\fourteenTypewriter}%
\textfont\bffam=\fourteenSansSerif%
\def\bf{\fam\bffam\fourteenSansSerif}\scriptfont\bffam=\tenSansSerif\scriptscriptfont\bffam=\eightSansSerif%
\def\cal{\fourteenSymbols}%
\def\greekbold{\fourteenMathBold}%
\def\gothic{\fourteenGothic}%
\def\Bbb{\fourteenDouble}%
\def\LieFont{\fourteenSerifItalics}%
\nt\normalbaselines\baselineskip=15pt%
}

\def\PartStyle{\parindent=0pt\parskip=0pt\normalbaselineskip=15pt%
\def\nt{\fourteenSansSerifBold}%
\def\rm{\fam0\fourteenSansSerifBold}%
\textfont0=\fourteenSansSerifBold\scriptfont0=\tenSansSerifBold\scriptscriptfont0=\eightSansSerifBold
\textfont1=\fourteenMath\scriptfont1=\tenMath\scriptscriptfont1=\eightMath
\textfont2=\fourteenSymbols\scriptfont2=\tenSymbols\scriptscriptfont2=\eightSymbols
\textfont3=\fourteenMoreSymbols\scriptfont3=\tenMoreSymbols\scriptscriptfont3=\eightMoreSymbols
\textfont\itfam=\fourteenSansSerifItalics\def\it{\fam\itfam\fourteenSansSerifItalics}%
\textfont\slfam=\fourteenSansSerifSlanted\def\sl{\fam\slfam\fourteenSerifSansSlanted}%
\textfont\ttfam=\fourteenTypewriter\def\tt{\fam\ttfam\fourteenTypewriter}%
\textfont\bffam=\fourteenSansSerif%
\def\bf{\fam\bffam\fourteenSansSerif}\scriptfont\bffam=\tenSansSerif\scriptscriptfont\bffam=\eightSansSerif%
\def\cal{\fourteenSymbols}%
\def\greekbold{\fourteenMathBold}%
\def\gothic{\fourteenGothic}%
\def\Bbb{\fourteenDouble}%
\def\LieFont{\fourteenSerifItalics}%
\nt\normalbaselines\baselineskip=15pt%
}

\def\ChapterStyle{\parindent=0pt\parskip=0pt\normalbaselineskip=15pt%
\def\nt{\fourteenSansSerifBold}%
\def\rm{\fam0\fourteenSansSerifBold}%
\textfont0=\fourteenSansSerifBold\scriptfont0=\tenSansSerifBold\scriptscriptfont0=\eightSansSerifBold
\textfont1=\fourteenMath\scriptfont1=\tenMath\scriptscriptfont1=\eightMath
\textfont2=\fourteenSymbols\scriptfont2=\tenSymbols\scriptscriptfont2=\eightSymbols
\textfont3=\fourteenMoreSymbols\scriptfont3=\tenMoreSymbols\scriptscriptfont3=\eightMoreSymbols
\textfont\itfam=\fourteenSansSerifItalics\def\it{\fam\itfam\fourteenSansSerifItalics}%
\textfont\slfam=\fourteenSansSerifSlanted\def\sl{\fam\slfam\fourteenSerifSansSlanted}%
\textfont\ttfam=\fourteenTypewriter\def\tt{\fam\ttfam\fourteenTypewriter}%
\textfont\bffam=\fourteenSansSerif%
\def\bf{\fam\bffam\fourteenSansSerif}\scriptfont\bffam=\tenSansSerif\scriptscriptfont\bffam=\eightSansSerif%
\def\cal{\fourteenSymbols}%
\def\greekbold{\fourteenMathBold}%
\def\gothic{\fourteenGothic}%
\def\Bbb{\fourteenDouble}%
\def\LieFont{\fourteenSerifItalics}%
\nt\normalbaselines\baselineskip=15pt%
}

\def\SectionStyle{\parindent=0pt\parskip=0pt\normalbaselineskip=13pt%
\def\nt{\twelveSansSerifBold}%
\def\rm{\fam0\twelveSansSerifBold}%
\textfont0=\twelveSansSerifBold\scriptfont0=\eightSansSerifBold\scriptscriptfont0=\eightSansSerifBold
\textfont1=\twelveMath\scriptfont1=\eightMath\scriptscriptfont1=\eightMath
\textfont2=\twelveSymbols\scriptfont2=\eightSymbols\scriptscriptfont2=\eightSymbols
\textfont3=\twelveMoreSymbols\scriptfont3=\eightMoreSymbols\scriptscriptfont3=\eightMoreSymbols
\textfont\itfam=\twelveSansSerifItalics\def\it{\fam\itfam\twelveSansSerifItalics}%
\textfont\slfam=\twelveSansSerifSlanted\def\sl{\fam\slfam\twelveSerifSansSlanted}%
\textfont\ttfam=\twelveTypewriter\def\tt{\fam\ttfam\twelveTypewriter}%
\textfont\bffam=\twelveSansSerif%
\def\bf{\fam\bffam\twelveSansSerif}\scriptfont\bffam=\eightSansSerif\scriptscriptfont\bffam=\eightSansSerif%
\def\cal{\twelveSymbols}%
\def\bg{\twelveMathBold}%
\def\gothic{\twelveGothic}%
\def\Bbb{\twelveDouble}%
\def\LieFont{\twelveSerifItalics}%
\nt\normalbaselines\baselineskip=13pt%
}

\def\SubSectionStyle{\parindent=0pt\parskip=0pt\normalbaselineskip=13pt%
\def\nt{\twelveSansSerifItalics}%
\def\rm{\fam0\twelveSansSerifItalics}%
\textfont0=\twelveSansSerifItalics\scriptfont0=\eightSansSerifItalics\scriptscriptfont0=\eightSansSerifItalics%
\textfont1=\twelveMath\scriptfont1=\eightMath\scriptscriptfont1=\eightMath%
\textfont2=\twelveSymbols\scriptfont2=\eightSymbols\scriptscriptfont2=\eightSymbols%
\textfont3=\twelveMoreSymbols\scriptfont3=\eightMoreSymbols\scriptscriptfont3=\eightMoreSymbols%
\textfont\itfam=\twelveSansSerif\def\it{\fam\itfam\twelveSansSerif}%
\textfont\slfam=\twelveSansSerifSlanted\def\sl{\fam\slfam\twelveSerifSansSlanted}%
\textfont\ttfam=\twelveTypewriter\def\tt{\fam\ttfam\twelveTypewriter}%
\textfont\bffam=\twelveSansSerifBold%
\def\bf{\fam\bffam\twelveSansSerifBold}\scriptfont\bffam=\eightSansSerifBold\scriptscriptfont\bffam=\eightSansSerifBold%
\def\cal{\twelveSymbols}%
\def\greekbold{\twelveMathBold}%
\def\gothic{\twelveGothic}%
\def\Bbb{\twelveDouble}%
\def\LieFont{\twelveSerifItalics}%
\nt\normalbaselines\baselineskip=13pt%
}

\def\AuthorStyle{\parindent=0pt\parskip=0pt\normalbaselineskip=14pt%
\def\nt{\tenSerif}%
\def\rm{\fam0\tenSerif}%
\textfont0=\tenSerif\scriptfont0=\sevenSerif\scriptscriptfont0=\fiveSerif
\textfont1=\tenMath\scriptfont1=\sevenMath\scriptscriptfont1=\fiveMath
\textfont2=\tenSymbols\scriptfont2=\sevenSymbols\scriptscriptfont2=\fiveSymbols
\textfont3=\tenMoreSymbols\scriptfont3=\sevenMoreSymbols\scriptscriptfont3=\fiveMoreSymbols
\textfont\itfam=\tenSerifItalics\def\it{\fam\itfam\tenSerifItalics}%
\textfont\slfam=\tenSerifSlanted\def\sl{\fam\slfam\tenSerifSlanted}%
\textfont\ttfam=\tenTypewriter\def\tt{\fam\ttfam\tenTypewriter}%
\textfont\bffam=\tenSerifBold%
\def\bf{\fam\bffam\tenSerifBold}\scriptfont\bffam=\sevenSerifBold\scriptscriptfont\bffam=\fiveSerifBold%
\def\cal{\tenSymbols}%
\def\greekbold{\tenMathBold}%
\def\gothic{\tenGothic}%
\def\Bbb{\tenDouble}%
\def\LieFont{\tenSerifItalics}%
\nt\normalbaselines\baselineskip=14pt%
}

\def\AddressStyle{\parindent=0pt\parskip=0pt\normalbaselineskip=14pt%
\def\nt{\eightSerif}%
\def\rm{\fam0\eightSerif}%
\textfont0=\eightSerif\scriptfont0=\sevenSerif\scriptscriptfont0=\fiveSerif
\textfont1=\eightMath\scriptfont1=\sevenMath\scriptscriptfont1=\fiveMath
\textfont2=\eightSymbols\scriptfont2=\sevenSymbols\scriptscriptfont2=\fiveSymbols
\textfont3=\eightMoreSymbols\scriptfont3=\sevenMoreSymbols\scriptscriptfont3=\fiveMoreSymbols
\textfont\itfam=\eightSerifItalics\def\it{\fam\itfam\eightSerifItalics}%
\textfont\slfam=\eightSerifSlanted\def\sl{\fam\slfam\eightSerifSlanted}%
\textfont\ttfam=\eightTypewriter\def\tt{\fam\ttfam\eightTypewriter}%
\textfont\bffam=\eightSerifBold%
\def\bf{\fam\bffam\eightSerifBold}\scriptfont\bffam=\sevenSerifBold\scriptscriptfont\bffam=\fiveSerifBold%
\def\cal{\eightSymbols}%
\def\greekbold{\eightMathBold}%
\def\gothic{\eightGothic}%
\def\Bbb{\eightDouble}%
\def\LieFont{\eightSerifItalics}%
\nt\normalbaselines\baselineskip=14pt%
}

\def\AbstractStyle{\parindent=0pt\parskip=0pt\normalbaselineskip=12pt%
\def\nt{\eightSerif}%
\def\rm{\fam0\eightSerif}%
\textfont0=\eightSerif\scriptfont0=\sevenSerif\scriptscriptfont0=\fiveSerif
\textfont1=\eightMath\scriptfont1=\sevenMath\scriptscriptfont1=\fiveMath
\textfont2=\eightSymbols\scriptfont2=\sevenSymbols\scriptscriptfont2=\fiveSymbols
\textfont3=\eightMoreSymbols\scriptfont3=\sevenMoreSymbols\scriptscriptfont3=\fiveMoreSymbols
\textfont\itfam=\eightSerifItalics\def\it{\fam\itfam\eightSerifItalics}%
\textfont\slfam=\eightSerifSlanted\def\sl{\fam\slfam\eightSerifSlanted}%
\textfont\ttfam=\eightTypewriter\def\tt{\fam\ttfam\eightTypewriter}%
\textfont\bffam=\eightSerifBold%
\def\bf{\fam\bffam\eightSerifBold}\scriptfont\bffam=\sevenSerifBold\scriptscriptfont\bffam=\fiveSerifBold%
\def\cal{\eightSymbols}%
\def\greekbold{\eightMathBold}%
\def\gothic{\eightGothic}%
\def\Bbb{\eightDouble}%
\def\LieFont{\eightSerifItalics}%
\nt\normalbaselines\baselineskip=12pt%
}

\def\RefsStyle{\parindent=0pt\parskip=0pt%
\def\nt{\eightSerif}%
\def\rm{\fam0\eightSerif}%
\textfont0=\eightSerif\scriptfont0=\sevenSerif\scriptscriptfont0=\fiveSerif
\textfont1=\eightMath\scriptfont1=\sevenMath\scriptscriptfont1=\fiveMath
\textfont2=\eightSymbols\scriptfont2=\sevenSymbols\scriptscriptfont2=\fiveSymbols
\textfont3=\eightMoreSymbols\scriptfont3=\sevenMoreSymbols\scriptscriptfont3=\fiveMoreSymbols
\textfont\itfam=\eightSerifItalics\def\it{\fam\itfam\eightSerifItalics}%
\textfont\slfam=\eightSerifSlanted\def\sl{\fam\slfam\eightSerifSlanted}%
\textfont\ttfam=\eightTypewriter\def\tt{\fam\ttfam\eightTypewriter}%
\textfont\bffam=\eightSerifBold%
\def\bf{\fam\bffam\eightSerifBold}\scriptfont\bffam=\sevenSerifBold\scriptscriptfont\bffam=\fiveSerifBold%
\def\cal{\eightSymbols}%
\def\greekbold{\eightMathBold}%
\def\gothic{\eightGothic}%
\def\Bbb{\eightDouble}%
\def\LieFont{\eightSerifItalics}%
\nt\normalbaselines\baselineskip=10pt%
}



%
%


\def\ModeYes{yes}
\def\ModeNo{no}

\def\ModeUndef{undefined}


\def\nx{\noexpand}
\def\ni{\noindent}
\def\newpage{\vfill\eject}

\def\ss{\vskip 5pt}
\def\ms{\vskip 10pt}
\def\bs{\vskip 20pt}

 \def\,{\mskip\thinmuskip}
 \def\!{\mskip-\thinmuskip}
 \def\>{\mskip\medmuskip}
 \def\;{\mskip\thickmuskip}

%
% ... 
%

\def\refsModePost{post}
\def\refsModeAuto{auto}

\def\dbRefsSatusModeOk{ok}
\def\dbRefsSatusModeError{error}
\def\dbRefsSatusModeWarning{warning}


\newcount\BNUM
\BNUM=0

\def\refs{}

\def\SetModePost{\xdef\refsMode{\refsModePost}}			
\SetModePost

\def\dbRefsStatusOk{%
	\xdef\dbRefsStatus{\dbRefsSatusModeOk}%
	\xdef\dbRefsError{\ModeNo}%
	\xdef\dbRefsWarning{\ModeNo}%
	\xdef\dbRefsInfo{\ModeNo}%
}

\def\dbRefs{%
}

\def\dbRefsGet#1{%
	\xdef\found{N}\xdef\ikey{#1}\dbRefsStatusOk%
	\xdef\key{\ModeUndef}\xdef\tag{\ModeUndef}\xdef\tail{\ModeUndef}%
	\dbRefs%
}

\def\NextRefsTag{%
	\global\advance\BNUM by 1%
}
\def\ShowTag#1{{\bf [#1]}}

\def\dbRefsInsert#1#2{%
\dbRefsGet{#1}%
\if\found Y %
   \xdef\dbRefsStatus{\dbRefsSatusModeWarning}%
   \xdef\dbRefsWarning{record is already there}%
   \xdef\dbRefsInfo{record not inserted}%
\else%
   \toks2=\expandafter{\dbRefs}%
   \ifx\refsMode\refsModeAuto \NextRefsTag
    \xdef\dbRefs{%
   	\the\toks2 \nx\xdef\nx\dbx{#1}%
	\nx\ifx\nx\ikey %
		\nx\dbx\nx\xdef\nx\found{Y}%
		\nx\xdef\nx\key{#1}%
		\nx\xdef\nx\tag{\the\BNUM}%
		\nx\xdef\nx\tail{#2}%
	\nx\fi}%
	\global\xdef\refs{\refs \ss\ni[\the\BNUM]\ #2\par}
   \fi%
   \ifx\refsMode\refsModePost 
    \xdef\dbRefs{%
   	\the\toks2 \nx\xdef\nx\dbx{#1}%
	\nx\ifx\nx\ikey %
		\nx\dbx\nx\xdef\nx\found{Y}%
		\nx\xdef\nx\key{#1}%
		\nx\xdef\nx\tag{\ModeUndef}%
		\nx\xdef\nx\tail{#2}%
	\nx\fi}%
   \fi%
\fi%
}

\def\dbRefsEdit#1#2#3{\dbRefsGet{#1}%
\if\found N 
   \xdef\dbRefsStatus{\dbRefsSatusModeError}%
   \xdef\dbRefsError{record is not there}%
   \xdef\dbRefsInfo{record not edited}%
\else%
   \toks2=\expandafter{\dbRefs}%
   \xdef\dbRefs{\the\toks2%
   \nx\xdef\nx\dbx{#1}%
   \nx\ifx\nx\ikey\nx\dbx %
	\nx\xdef\nx\found{Y}%
	\nx\xdef\nx\key{#1}%
	\nx\xdef\nx\tag{#2}%
	\nx\xdef\nx\tail{#3}%
   \nx\fi}%
\fi%
}

\def\bib#1#2{\RefsStyle\dbRefsInsert{#1}{#2}%
	\ifx\dbRefsStatus\dbRefsSatusModeWarning %
		\message{^^J}%
		\message{WARNING: Reference [#1] is doubled.^^J}%
	\fi%
}

\def\ref#1{\dbRefsGet{#1}%
\ifx\found N %
  \message{^^J}%
  \message{ERROR: Reference [#1] unknown.^^J}%
  \ShowTag{??}%
\else%
	\ifx\tag\ModeUndef \NextRefsTag%
		\dbRefsEdit{#1}{\the\BNUM}{\tail}%
		\dbRefsGet{#1}%
		\global\xdef\refs{\refs \ss\ni [\tag]\ \tail\par}
	\fi
	\ShowTag{\tag}%
\fi%
}

\def\ShowBiblio{\ms\Ensure{\SectionEnsure}%
{\SectionStyle\ni References}%
{\RefsStyle\refs}%
}

\newcount\CHANGES
\CHANGES=0
\def\AuxFile{7}
\def\PreventDoubleOn{\xdef\PreventDoubleLabel{\ModeYes}}

\PreventDoubleOn

\def\StoreLabel#1#2{\xdef\itag{#2}
 \ifx\PreModeStatus\ModeNo %
   \message{^^J}%
   \errmessage{You can't use Check without starting with OpenPreMode (and finishing with ClosePreMode)^^J}%
 \else%
   \immediate\write\AuxFile{\nx\dbLabelPreInsert{#1}{\itag}}%
   \dbLabelGet{#1}%
   \ifx\itag\tag %
   \else%
	\global\advance\CHANGES by 1%
 	\xdef\itag{(?.??)}%
    \fi%
   \fi%
}

\def\PreModeStatus{\ModeNo}

\def\ClosePreMode{\immediate\closeout\AuxFile%
  \ifnum\CHANGES=0%
	\message{^^J}%
	\message{**********************************^^J}%
	\message{**  NO CHANGES TO THE AuxFile  **^^J}%
	\message{**********************************^^J}%
 \else%
	\message{^^J}%
	\message{**************************************************^^J}%
	\message{**  PLAEASE TYPESET IT AGAIN (\the\CHANGES)  **^^J}%
    \errmessage{**************************************************^^ J}%
  \fi%
  \edef\PreModeStatus{\ModeNo}%
}

\def\dbLabelSatusModeOk{ok}

\def\dbLabelSatusModeWarning{warning}

\def\dbLabelStatusOk{%
	\xdef\dbLabelStatus{\dbLabelSatusModeOk}%
	\xdef\dbLabelError{\ModeNo}%
	\xdef\dbLabelWarning{\ModeNo}%
	\xdef\dbLabelInfo{\ModeNo}%
}

\def\dbLabel{%
}

\def\dbLabelGet#1{%
	\xdef\found{N}\xdef\ikey{#1}\dbLabelStatusOk%
	\xdef\key{\ModeUndef}\xdef\tag{\ModeUndef}\xdef\pre{\ModeUndef}%
	\dbLabel%
}

\def\ShowLabel#1{%
 \dbLabelGet{#1}%
 \ifx\tag \ModeUndef %
 	\global\advance\CHANGES by 1%
 	(?.??)%
 \else%
 	\tag%
 \fi%
}

\def\dbLabelPreInsert#1#2{\dbLabelGet{#1}%
\if\found Y %
  \xdef\dbLabelStatus{\dbLabelSatusModeWarning}%
   \xdef\dbLabelWarning{Label is already there}%
   \xdef\dbLabelInfo{Label not inserted}%
   \message{^^J}%
   \errmessage{Double pre definition of label [#1]^^J}%
\else%
   \toks2=\expandafter{\dbLabel}%
    \xdef\dbLabel{%
   	\the\toks2 \nx\xdef\nx\dbx{#1}%
	\nx\ifx\nx\ikey %
		\nx\dbx\nx\xdef\nx\found{Y}%
		\nx\xdef\nx\key{#1}%
		\nx\xdef\nx\tag{#2}%
		\nx\xdef\nx\pre{\ModeYes}%
	\nx\fi}%
\fi%
}

\def\dbLabelInsert#1#2{\dbLabelGet{#1}%
\xdef\itag{#2}%
\dbLabelGet{#1}%
\if\found Y %
	\ifx\tag\itag %
	\else%
	   \ifx\PreventDoubleLabel\ModeYes %
		\message{^^J}%
		\errmessage{Double definition of label [#1]^^J}%
	   \else%
		\message{^^J}%
		\message{Double definition of label [#1]^^J}%
	   \fi%
	\fi%
   \xdef\dbLabelStatus{\dbLabelSatusModeWarning}%
   \xdef\dbLabelWarning{Label is already there}%
   \xdef\dbLabelInfo{Label not inserted}%
\else%
   \toks2=\expandafter{\dbLabel}%
    \xdef\dbLabel{%
   	\the\toks2 \nx\xdef\nx\dbx{#1}%
	\nx\ifx\nx\ikey %
		\nx\dbx\nx\xdef\nx\found{Y}%
		\nx\xdef\nx\key{#1}%
		\nx\xdef\nx\tag{#2}%
		\nx\xdef\nx\pre{\ModeNo}%
	\nx\fi}%
\fi%
}


\newcount\PART
\newcount\CHAPTER
\newcount\SECTION
\newcount\SUBSECTION
\newcount\FNUMBER

\PART=0
\CHAPTER=0
\SECTION=0
\SUBSECTION=0	
\FNUMBER=0

\def\LastPart{\ModeUndef}
\def\LastChapter{\ModeUndef}
\def\LastSection{\ModeUndef}
\def\LastSubSection{\ModeUndef}
\def\LastClaim{\ModeUndef}
\def\Last{\ModeUndef}

\newdimen\TOBOTTOM
\newdimen\LIMIT

\def\Ensure#1{\ \par\ \immediate\LIMIT=#1\immediate\TOBOTTOM=\the\pagegoal\advance\TOBOTTOM by -\pagetotal%
\ifdim\TOBOTTOM<\LIMIT\newpage \else%
\vskip-\parskip\vskip-\parskip\vskip-\baselineskip\fi}

\def\PartLabel{\the\PART}
\def\NewPart#1{\global\advance\PART by 1%
         \bs\ni{\PartStyle  Part \PartLabel:}
         \bs\ni{\PartStyle #1}\newpage%
         \CHAPTER=0\SECTION=0\SUBSECTION=0\FNUMBER=0%
         \gdef\Left{#1}%
         \global\edef\Last{\PartLabel}%
         \global\edef\LastPart{\PartLabel}%
         \global\edef\LastChapter{\ModeUndef}%
         \global\edef\LastSection{\ModeUndef}%
         \global\edef\LastSubSection{\ModeUndef}%
         \global\edef\LastClaim{\ModeUndef}}
\def\ChapterLabel{\the\CHAPTER}
\def\NewChapter#1{\global\advance\CHAPTER by 1%
         \bs\ni{\ChapterStyle  Chapter \ChapterLabel: #1}\ms%
         \SECTION=0\SUBSECTION=0\FNUMBER=0%
         \gdef\Left{#1}%
         \global\edef\Last{\ChapterLabel}%
         \global\edef\LastChapter{\ChapterLabel}%
         \global\edef\LastSection{\ModeUndef}%
         \global\edef\LastSubSection{\ModeUndef}%
         \global\edef\LastClaim{\ModeUndef}}
\def\SectionEnsure{3cm}
\def\NewSection#1{\Ensure{\SectionEnsure}\gdef\SectionLabel{\the\SECTION}\global\advance\SECTION by 1%
         \ms\ni{\SectionStyle  \SectionLabel.\ #1}\ss%
         \SUBSECTION=0\FNUMBER=0%
         \gdef\Left{#1}%
         \global\edef\Last{\SectionLabel}%
         \global\edef\LastSection{\SectionLabel}%
         \global\edef\LastSubSection{\ModeUndef}%
         \global\edef\LastClaim{\ModeUndef}}
\def\NewAppendix#1#2{\Ensure{\SectionEnsure}\gdef\SectionLabel{#1}\global\advance\SECTION by 1%
         \bs\ni{\SectionStyle  Appendix \SectionLabel.\ #2}\ss%
         \SUBSECTION=0\FNUMBER=0%
         \gdef\Left{#2}%
         \global\edef\Last{\SectionLabel}%
         \global\edef\LastSection{\SectionLabel}%
         \global\edef\LastSubSection{\ModeUndef}%
         \global\edef\LastClaim{\ModeUndef}}
\def\Acknowledgements{\Ensure{\SectionEnsure}\gdef\SectionLabel{}%
         \ms\ni{\SectionStyle  Acknowledgments}\ss%
         \SECTION=0\SUBSECTION=0\FNUMBER=0%
         \gdef\Left{}%
         \global\edef\Last{\ModeUndef}%
         \global\edef\LastSection{\ModeUndef}%
         \global\edef\LastSubSection{\ModeUndef}%
         \global\edef\LastClaim{\ModeUndef}}
\def\SubSectionEnsure{2cm}
\def\SubSectionLabel{\ifnum\SECTION>0 \the\SECTION.\fi\the\SUBSECTION}
\def\NewSubSection#1{\Ensure{\SubSectionEnsure}\global\advance\SUBSECTION by 1%
         \ms\ni{\SubSectionStyle #1}\ss%
         \global\edef\Last{\SubSectionLabel}%
         \global\edef\LastSubSection{\SubSectionLabel}}
\def\SetNumberingModeN{\def\ClaimLabel{(\the\FNUMBER)}}
\def\SetNumberingModeSN{\def\ClaimLabel{(\ifnum\SECTION>0 \SectionLabel.\fi%
      \the\FNUMBER)}}
\def\SetNumberingModeCSN{\def\ClaimLabel{(\ifnum\CHAPTER>0 \the\CHAPTER.\fi%
      \ifnum\SECTION>0 \SectionLabel.\fi%
      \the\FNUMBER)}}

\def\NewClaim{\global\advance\FNUMBER by 1%
    \ClaimLabel%
    \global\edef\LastClaim{\ClaimLabel}%
    \global\edef\Last{\ClaimLabel}}

\def\HideLabels{\xdef\ShowLabelsMode{\ModeNo}}
\HideLabels

\def\fn{\eqno{\NewClaim}} 
\def\fl#1{%
\ifx\ShowLabelsMode\ModeYes%
 \eqno{{\buildrel{\hbox{\AbstractStyle[#1]}}\over{\hfill\NewClaim}}}%
\else%
 \eqno{\NewClaim}%
\fi%
\dbLabelInsert{#1}{\ClaimLabel}}
\def\fprel#1{\global\advance\FNUMBER by 1\StoreLabel{#1}{\ClaimLabel}%
\ifx\ShowLabelsMode\ModeYes%
\eqno{{\buildrel{\hbox{\AbstractStyle[#1]}}\over{\hfill.\itag}}}%
\else%
 \eqno{\itag}%
\fi%
}

\def\cl#1{\global\advance\FNUMBER by 1\dbLabelInsert{#1}{\ClaimLabel}%
\ifx\ShowLabelsMode\ModeYes%
${\buildrel{\hbox{\AbstractStyle[#1]}}\over{\hfill\ClaimLabel}}$%
\else%
  $\ClaimLabel$%
\fi%
}
\def\cprel#1{\global\advance\FNUMBER by 1\StoreLabel{#1}{\ClaimLabel}%
\ifx\ShowLabelsMode\ModeYes%
${\buildrel{\hbox{\AbstractStyle[#1]}}\over{\hfill.\itag}}$%
\else%
  $\itag$%
\fi%
}

\def\Note{\ms\leftskip 3cm\rightskip 1.5cm\AbstractStyle}
\def\endNote{\par\leftskip 2cm\rightskip 0cm\NormalStyle\ss}


\parindent=7pt
\leftskip=2cm
\newcount\SideIndent
\newcount\SideIndentTemp
\SideIndent=0
\newdimen\SectionIndent
\SectionIndent=-8pt

\def\sidebar{\vrule height15pt width.2pt }
\def\endcorner{\hbox{\hbox{\vrule height6pt width.2pt}\vbox to6pt{\vfill\hbox
to4pt{\leaders\hrule height0.2pt\hfill}}}}
\def\begincorner{\hbox{\hbox{\vrule height6pt width.2pt}\vbox to6pt{\hbox
to4pt{\leaders\hrule height0.2pt\hfill}}}}
\def\endbegincorner{\hbox{\vbox to15pt{\endcorner\vskip-6pt\begincorner\vfill}}}
\def\SideShow{\SideIndentTemp=\SideIndent \ifnum \SideIndentTemp>0 
\loop\sidebar\hskip 2pt \advance\SideIndentTemp by-1\ifnum \SideIndentTemp>1 \repeat\fi}

\def\BeginSection{{\vbadness 100000 \par\ni\hskip\SectionIndent%
\SideShow\vbox to 15pt{\vfill\begincorner}}\global\advance\SideIndent by1\vskip-10pt}

\def\EndSection{{\vbadness 100000 \par\ni\global\advance\SideIndent by-1%
\hskip\SectionIndent\SideShow\vbox to15pt{\endcorner\vfill}\vskip-10pt}}

\def\EndBeginSection{{\vbadness 100000\par\ni%
\global\advance\SideIndent by-1\hskip\SectionIndent\SideShow
\vbox to15pt{\vfill\endbegincorner}}%
\global\advance\SideIndent by1\vskip-10pt}

\def\ShowBeginCorners#1{%
\SideIndentTemp =#1 \advance\SideIndentTemp by-1%
\ifnum \SideIndentTemp>0 %
\vskip-15truept\hbox{\kern 2truept\vbox{\hbox{\begincorner}%
\ShowBeginCorners{\SideIndentTemp}\vskip-3truept}}%
\fi%
}

\def\ShowEndCorners#1{%
\SideIndentTemp =#1 \advance\SideIndentTemp by-1%
\ifnum \SideIndentTemp>0 %
\vskip-15truept\hbox{\kern 2truept\vbox{\hbox{\endcorner}%
\ShowEndCorners{\SideIndentTemp}\vskip 2truept}}%
\fi%
}

\def\BeginSections#1{{\vbadness 100000 \par\ni\hskip\SectionIndent%
\SideShow\vbox to 15pt{\vfill\ShowBeginCorners{#1}}}\global\advance\SideIndent by#1\vskip-10pt}

\def\EndSections#1{{\vbadness 100000 \par\ni\global\advance\SideIndent by-#1%
\hskip\SectionIndent\SideShow\vbox to15pt{\vskip15pt\ShowEndCorners{#1}\vfill}\vskip-10pt}}

\def\EndBeginSections#1#2{{\vbadness 100000\par\ni%
\global\advance\SideIndent by-#1%
\hbox{\hskip\SectionIndent\SideShow\kern-2pt%
\vbox to15pt{\vskip15pt\ShowEndCorners{#1}\vskip4pt\ShowBeginCorners{#2}}}}%
\global\advance\SideIndent by#2\vskip-10pt}




%
%


\def\al{\alpha}
\def\be{\beta}
\def\de{\delta}
\def\ga{\gamma}

\def\ep{\epsilon}

\def\la{\lambda}

\def\om{\omega}

\def\vp{\varphi}

\def\Ga{\Gamma}

\def\Si{\Sigma}



 \def\gotP{{\hbox{\gothic P}}}
 


 \def\R{{\hbox{\Bbb R}}}

 \def\R{{\hbox{\Bbb R}}}


\def\ip{\hbox to4pt{\leaders\hrule height0.3pt\hfill}\vbox to8pt{\leaders\vrule width0.3pt\vfill}\kern 2pt}
 
\def\del{\partial}
\def\na{\nabla}

\def\Lie{\hbox{\LieFont \$}}

\def\arr{\rightarrow}

\def\then{\Rightarrow}

%
%

\def\cases#1{\left\{\eqalign{#1}\right.}
\NormalStyle
\SetNumberingModeSN
\PreventDoubleOn

\long\def\title#1{\centerline{\TitleStyle\ni#1}}

\long\def\author#1{\ms\centerline{\AuthorStyle by {\it #1}}}

\long\def\address#1{\ss\centerline{\AddressStyle #1}\par}
\long\def\moreaddress#1{\centerline{\AddressStyle #1}\par}
\def\abstract{\ms\leftskip 3cm\rightskip .5cm\AbstractStyle{\bf \ni Abstract:}\ }
\def\endabstract{\par\leftskip 2cm\rightskip 0cm\NormalStyle\ss}

\SetNumberingModeSN

\def\nab#1{{\buildrel #1\over \na}}
\def\frac[#1/#2]{\hbox{$#1\over#2$}}
\def\Frac[#1/#2]{{#1\over#2}}
\def\({\left(}
\def\){\right)}
\def\[{\left[}
\def\]{\right]}
\def\^#1{{}^{#1}_{\>\cdot}}
\def\_#1{{}_{#1}^{\>\cdot}}
\def\Label=#1{{\buildrel {\hbox{\fiveSerif \ShowLabel{#1}}}\over =}}
\def\<{\kern -1pt}


\def\CollapseAllCNotes{\long\def\CNote##1{}}
\def\ExpandAllCNotes{\long\def\CNote##1{%
\BeginSection
	\Note%
 		##1%
	\endNote%
\EndSection%
}}
\ExpandAllCNotes
%
%
%
%


\def\red#1{\textcolor{red}{#1}}
\def\blue#1{\textcolor{blue}{#1}}

\def\frame#1{\vbox{\hrule\hbox{\vrule\vbox{\kern2pt\hbox{\kern2pt#1\kern2pt}\kern2pt}\vrule}\hrule\kern-4pt}}

\def\uline#1{\underline{#1}}

\def\Box to #1#2#3{\frame{\vtop{\hbox to #1{\hfill #2 \hfill}\hbox to #1{\hfill #3 \hfill}}}}


\bib{EPS}{J.Ehlers, F.A.E.Pirani, A.Schild, 
{\it The Geometry of Free Fall and Light Propagation},
in General Relativity, ed. L.OÕRaifeartaigh (Clarendon, Oxford, 1972). 
}

\bib{EPS1}
{M.Di Mauro, L. Fatibene, M.Ferraris, M.Francaviglia, 
{\it Further Extended Theories of Gravitation: Part I },
Int. J. Geom. Methods Mod. Phys. Volume: 7, Issue: 5 (2010), pp. 887-898; gr-qc/0911.2841}

\bib{EPS2}
{L. Fatibene, M.Ferraris, M.Francaviglia, S.Mercadante,
{\it Further Extended Theories of Gravitation: Part II},
Int. J. Geom. Methods Mod. Phys. Volume: 7, Issue: 5 (2010), pp. 899-906; gr-qc/0911.284}

\bib{ELQG}
{L. Fatibene, M. Ferraris, M. Francaviglia,
{\it Extended Loop Quantum Gravity},
CQG 27(18) 185016 (2010); arXiv:1003.1619}

\bib{MGaCou}
{L.Fatibene, M.Francaviglia, S. Mercadante,
{\it Matter Lagrangians Coupled with Connections}
Int. J. Geom. Methods Mod. Phys. Volume: 7, Issue: 5 (2010), 1185-1189; arXiv: 0911.2981}

\bib{catalogo}{ H.Stephani, D.Kramer, M.Mac Callum,
{\it Exact Solutions Of Einstein's Field Equations},
Cambridge University Press (2003) }

\bib{Perlick}{V. Perlick, 
{\it Characterization of standard clocks by means of light rays and freely falling particles}
General Relativity and Gravitation,  {\bf 19}(11) (1987) 1059-1073}

\bib{Faraoni}{T.P. Sotiriou, V. Faraoni,
{\it  $f (R)$  theories of gravity}, (2008); 
arXiv: 0805.1726v2
}

\bib{Capozziello}{S. Capozziello, M. De Laurentis, V. Faraoni
{\it A bird's eye view of $f(R)$-gravity}
(2009); arXiv:0909.4672 
}

\bib{Magnano}{G. Magnano, L.M. Sokolowski, 
{\it On Physical Equivalence between Nonlinear Gravity Theories}
Phys.Rev. D50 (1994) 5039-5059; gr-qc/9312008
}

\bib{S2}{T.P. Sotiriou,
{\it $f(R)$ gravity, torsion and non-metricity},
Class. Quant. Grav. 26 (2009) 152001; gr-qc/0904.2774}

\bib{S3}{T.P. Sotiriou,
{\it Modified Actions for Gravity: Theory and Phenomenology},
Ph.D. Thesis; gr-qc/0710.4438}

\bib{C1}{S. Capozziello, M. Francaviglia,
{\it Extended Theories of Gravity and their Cosmological and Astrophysical Applications},
Journal of General Relativity and Gravitation 40 (2-3), (2008) 357-420.}

\bib{C2}{S. Capozziello, M.F. De Laurentis, M. Francaviglia, S. Mercadante,
{\it From Dark Energy and Dark Matter to Dark Metric},
Foundations of Physics 39 (2009) 1161-1176
gr-qc/0805.3642v4}

\bib{C4}{S. Capozziello, M. De Laurentis, M. Francaviglia, S. Mercadante,
{\it First Order Extended Gravity and the Dark Side of the Universe Ð II: Matching Observational Data},
Proceedings of the Conference ``Univers Invisibile'', Paris June 29 Ð July 3, 2009 
Ð to appear in 2010}

\bib{Olmo}{Olmo Sigh}

\bib{Schouten}{
J.A.Schouten,
{\it Ricci-Calculus: An Introduction to Tensor Analysis and its Geometrical Applications},
Springer Verlag (1954)}

\bib{Bibliopolis}{M.Francaviglia,
{\it Relativistic theories},
Quaderni del CNR-GNFM (Italy, 1988)
}

\bib{Moon}{T.Y.Moon, J.Lee, P.Oh,
{\it Conformal invariance in Einstein-Cartan-Weyl space},
Mod. Phys. Lett. A25, 3129 (2010); arXiv:0912.0432[gr-qc]
}

\bib{UnyFerraris}{M.Ferraris, J.Kijowski, 
{\it Unified geometric theory of electromagnetic and gravitational interactions}
GRG {\bf 14}(1) (1982), 37-47
}

\bib{Fluidi1}{L. Fatibene, M. Francaviglia, 
{\it Weyl Geometries and Timelike Geodesics},
arXiv:1106.1961v1 [gr-qc]
}

\bib{Reina}{M. Ferraris, M. Francaviglia, C. Reina,
{\it Variational Formulation of General Relativity from 1915 to 1925. ``Palatini's Method'' Discovered by Einstein in 1925},
GRG{\bf 14}(3), 243-254 (1982)
}



\def\ubal{\underline{\al}\kern1pt}
\def\obal{\overline{\al}\kern1pt}

\def\ubR{\underline{R}\kern1pt}
\def\obR{\overline{R}\kern1pt}
\def\ubom{\underline{\om}\kern1pt}
\def\obxi{\overline{\xi}\kern1pt}
\def\ubu{\underline{u}\kern1pt}
\def\ube{\underline{e}\kern1pt}
\def\obe{\overline{e}\kern1pt}

\NormalStyle
\CollapseAllCNotes

\title{Fluids in Weyl Geometries}

\author{L.Fatibene$^{a,b}$, M.Francaviglia$^{a,b}$}

\address{$^a$ Department of Mathematics, University of Torino (Italy)}
\moreaddress{$^b$ INFN - Sez. Torino. Iniziativa Specifica Na12}

\abstract
We shall investigate the consequences of non-trivial Weyl geometries on conservation laws of a fluid.
In particular we shall obtain a set of properties which allow to obtain in this generalized setting the standard relation 
between conservation of the energy-momentum tensor and number of particles.
\endabstract

\NewSection{Introduction}

All physical experiments use electromagnetic field in quite a peculiar way. Although this issue was usually overlooked (and practically irrelevant) at the time of mechanical experiments, it was finally stressed during 19th and 20th centuries when Physics recognized to be working in terms of fields.

With special relativity (SR) Einstein realized that the physical properties of the electromagnetic field are fundamentally entangled with our notion of space and time (or, more correctly, {\it spacetime}) on which our observational protocols rely, for example through synchronization. 
SR is in fact a good model of spacetime and electromagnetic field in which observational protocols are made explicit. As such, and in view of later studies of general relativity (GR), SR is fundamentally a theory of empty space, with electromagnetic field but no other matter. 

As soon as massive particles are considered they generate a gravitational field that cannot be described in SR. Of course one can in some case neglect gravitational effects and consider massive particles as test particles; however, from a fundamental viewpoint SR is not compatible with masses.

When Einstein was led to consider the gravitational field he was forced to take the only direction that his contemporary geometric technologies allowed to use. Gravitation was identified with the curvature of a spacetime metric, leading to standard GR.  
At that time, general connections were still on the way of being invented and the only curvature known was metric curvature. 

When general connections were finally invented (maybe feeling something to be improved in GR in the relation among electromagnetic field, gravitational field and matter) Einstein tried the way of a unifying theory of electromagnetic and gravitational  fields in which both fields were the output of a connection.  Einstein could not reach a satisfactory solution to this problem (later solved in \ref{UnyFerraris}).

In 1972 another fundamental contribution was given by Ehlers, Pirani and Schild (EPS). They started by assuming that one can observe and trace lightrays and particle worldlines in spacetime and they listed a set of axioms about these congruences of trajectories together with their compatibility, resorting to physically reasonable properties. EPS axioms allow to derive the geometric structure of spacetime starting from potentially observable data about lightrays and particles. And the final conclusion of EPS is that lightrays determine a conformal structure on spacetime (i.e.\ a sheaf of lightcones or a class $[g]$ of conformally equivalent Lorentzian metrics) and massive particles move along geodesic worldlines of a connection $\Ga$ which is neither physically nor mathematically  forced to be the Levi-Civita connection of the metric determined by lightrays.
These two structures have just to be {\it EPS-compatible} in the sense that lightlike geodesics with respect to $\{g\}$ have to belong to the family of auto parallel curves of  $\Ga$. It can be shown that that happens iff the connection $\Ga$ is chosen so that there exists a covector $A$ such that:
$$
\Ga^\al_{\be\mu} = \{ g\}^\al_{\be\mu}  + (g^{\al\ep}g_{\be\mu} -2 \de^\al_{(\be}\de^\ep_{\mu)}) A_\ep
\fl{EPSCompatibility}$$

In other words, EPS showed that lightrays and mass particles uniquely determine a so-called {\it Weyl geometry} $(M, [g], \Gamma)$. In this setting the conformal structure $[g]$ is directly related to lightrays and the connection $\Ga$ is directly related to the free fall of massive particles.

EPS framework clearly shows that it is unlogical to require or impose at a purely kinematical level that the connection is the Levi-Civita connection of the metric $g$. There is no reason to assume such a strict relation between the metric and affine structures. Of course, being unlogical it does not imply being physically false. Standard GR, in which these two structures do in fact coincide by assumption, might be the {\it true} choice, though that should be decided by experiments and not assumed {\it a priori}. 

EPS framework clearly points towards a formalism {\it\`a la Palatini} (which by the way was another finding of Einstein; see \ref{Reina}), in which the metric and the connection are {\it a priori} unrelated and field equations then decide dynamically the relation between the two structures. It is well-known that in vacuum (or with electromagnetism and scalar matter fields) in standard Palatini formalism the connection is forced to be the Levi-Civita connection of the metric, which would in fact explain why GR works so well in simple situations such as the experiments in the Solar system.

By the way, there is another issue related to EPS formalism which is against the assumptions of standard GR. In EPS lightrays determine a conformal structure, not a metric. Choosing a representative of the conformal structure, i.e. a Lorentzian metric $g$, to represent the conformal structure is in fact a sort of fixing of the conformal gauge invariance which the dynamics of the gravitational field has to preserve. Hence one has conformal transformations acting on metrics and on connections and the Lagrangian has to be covariant with respect to such transformations. However, conformal transformations are related to lightrays, not to massive particles, and the connection $\Ga$ is left unchanged by such transformations. Accordingly, a conformal transformation has a particular physical meaning and it acts by rescaling the metric but leaving the connection unchanged.
If so, imposing kinematically a strict relation between the connection and the metric, namely $\Ga=\{g\}$, would be incompatible with such a group symmetry.

In any event, physically speaking one should resign to determine the metric structure by just observing lightrays (or more generally the electromagnetic field which in dimension $m=4$ is known to be conformally invariant). The metric struture is then historically used to gauge observational protocols and experiments.
Then one should observe massive particle free fall and determine the connection among the EPS-compatible structures. 
In other words, electromagnetic and mass phenomenologies are distinct both from the experimental point of view and on the role they play in the definition of the geometry of spacetime.

Now, we want to suggest that this new perspective opens the possibility to solve in a different more general way the problem 
faced by Einstein of finding a model of gravitational, electromagnetic and matter fields by assuming a much larger class of theories, in which standard GR is just one possibility, in which one should set up experiments to determine which is the theory which is physically realized by the gravity we know.
In particular we shall show how one could consider a matter fluid on Minkowski spacetime, generating a gravitational field which is encoded in a (curved) Weyl connection which is compatible with the flat Minkowski metric structure. This spacetime is curved, in the sense that the connection $\Ga$ {\it is curved}, even though the metric structure is the standard flat Minkowski which is in fact unaffected by the presence of the matter fluid, being it determined by the Physics of lightrays.

\ms
Hereafter, in Section 2 we shall briefly review how a fluid on a conformal structure $(M, [g])$ essentially determines
a (unique) Weyl geometry $(M, [g], \Ga)$ in which in fact the fluid flows along geodesics of $\Ga$; see \ref{Fluidi1}.

In Section 3 we shall investigate how the system behaves from the viewpoint of its stress tensors and its relation to conservation laws and geodesics of $\Ga$. In standard GR this issue is very well known. 
However, in Weyl  geometries one has two connections (namely, $\{g\}$ and $\Ga$) and whenever a connection is used one has to declare which one of these two has to be used. This cannot be done without constraints; the geodesic equation of $\Ga$ should be made 
to work together with conservation of the stress tensor (and Bianchi identities) in order to produce the expected conservation laws of the fluid. 

In Section 4 the relation between the conservation of the number of particles is related to the conservation of the energy-momentum tensor. We shall show that this can be done by selecting a preferred conformal frame and a parametrization of worldlines.

\NewSection{Weyl Geometries Compatible with a Fluid}

Hereafter  we shall briefly review the results of \ref{Fluidi1}.
Let us consider, on an spacetime $M$ of dimension $m$, a conformal structure $(M, \gotP)$ and any $\gotP$-timelike vector field $u$.
For any gauge fixing of the conformal structure $g\in \gotP\equiv [g]$ one can normalize $u$ to be a $g$-unit vector $n$.

Let $\ga:\R\arr M$ be an integral curve of the vector field $n$. 
We can arbitrarily reparametrize the curve $\ga$ to obtain a different representative $\ga\circ \phi$ of the same trajectory.
If the original curve $\ga$ was a $\Ga$-geodesic motion (for a connection $\Ga$) then $\ga\circ \phi$ is a $\Ga$-geodesic trajectory.
Accordingly, one has
$$
n^\mu \nab{\Ga}_\mu n^\al = \vp \cdot n^\al
\fl{GeodesicCondition}$$
for some scalar field $\vp(x)$.

In GR one has no much choice for the fluid flow lines generated by $n$; the connection is freezed to be the Levi-Civita connection of $g$ and the vector field $n$ has to be selected in the small class of geodesic fields.
In a Weyl setting one has  a wider freedom in chosing the connection in the class of EPS-compatible connections given by \ShowLabel{EPSCompatibility}.  One can rely on this freedom to show that for any timelike vector field $n$ there exists an
EPS-compatible connection $\Ga$ for which $n$ is $\Ga$-geodesic, i.e.~\ShowLabel{GeodesicCondition} holds true.

One can easily check that $A$ has to fixed as
$$
A_\nu = n^\mu \nab{g}_\mu n_\nu  + \vp n_\nu
\fl{FluidConnection}$$
Notice how, once $u$ is given and a parametrization of curves is fixed by choosing the scalar field $\vp$,
one can uniquely determine the covector $A$ and thence the connection $\Ga$.

If one started from a different conformal representative $\tilde g=\Phi^2 \cdot g$ this would amount to redefine 
accordingly 
the covector $\tilde A=  A +  d \ln \Phi$ to obtain the same $\Ga$,
the unit vector $\tilde n^\la= \frac[1/\Phi] n^\la$ and 
 the scalar field $\tilde \vp= \frac[1/\Phi]  \( \vp  -   n^\mu \nab{\ast}_\mu \ln\Phi  \)$,
where $\nab{\ast}_\mu$ denotes the covariant derivative in the case it is independent of the connection. 
We refer to \ref{Fluidi1} for details and proofs.

\

\NewSection{Fluid Conservation Laws}

The fluid is described, besides by its flow lines which are generated by $u$, by two scalar fields $\rho(x)$ and $p(x)$
describing particle density and pressure. The  (symmetric)  energy-momentum tensor of the fluid is in the form
$$
T_{\mu\nu} = p g_{\mu\nu} + (p+\rho ) n_\mu n_\nu
\fn$$

In standard GR, one has a strict relation between Bianchi identities (associated to field equations), conservation of 
energy-monentum tensor $\na_\nu T^{\mu\nu}=0$, matter field equations 
(which together with fluid equation of state determine the evolution of the fluid) and 
conservation laws which are associated to conservation of the number of particles (and the fluid energy).

In Weyl setting one should describe the same sort of relations, by suitably specifying when covariant derivatives are induced by $g$ and when they are induced by $\Ga$.

The conservation of energy-momentum tensor  is
$$
\nab{\Ga}_\nu T^{\mu\nu}=0
\quad\then\qquad
\cases{
&(p+\rho )  \nab{\Ga}_\nu n^\nu=(p-\rho)    \vp-  n^\nu \nab{\ast}_\nu \rho  \cr
&\(g^{\mu\nu}+ n^\mu n^\nu  \) \nab{\ast}_\nu p = 2p  n^\nu \nab{g}_\nu n^\mu \cr
}
\fn$$
where we used \ShowLabel{EPSCompatibility} and \ShowLabel{FluidConnection}.

\CNote{
Conservtion equation can be projected along $n$. 
$$
\eqalign{
\nab{\Ga}_\nu T^{\mu\nu} = &
\nab{\ast}_\nu p g^{\mu\nu} 
+ p  \nab{\Ga}_\nu g^{\mu\nu} 
+ \nab{\ast}_\nu (p+\rho ) n^\mu n^\nu
+(p+\rho ) n^\nu \nab{\Ga}_\nu n^\mu 
+(p+\rho ) n^\mu \nab{\Ga}_\nu n^\nu=\cr
=&\nab{\ast}_\nu p g^{\mu\nu} 
-2p  A_\nu  g^{\mu\nu} + 
\nab{\ast}_\nu (p+\rho ) n^\mu n^\nu 
+(p+\rho ) \vp  n^\mu 
+(p+\rho ) n^\mu \nab{\Ga}_\nu n^\nu\cr
}
\fl{ConservationT}$$

This is a vector equation and can be decomposed in the direction of $n$, i.e. 
$$
\eqalign{
n_\mu \nab{\Ga}_\nu T^{\mu\nu} = &
\red{n^\nu  \nab{\ast}_\nu p}
-2p  n^\mu  A_\nu  
-  \nab{\ast}_\nu (\red{p}+\rho ) n^\nu 
-p \vp   -\rho  \vp   
-(p+\rho )  \nab{\Ga}_\nu n^\nu=\cr
 = &
-\frame{$2p   n^\mu n^\nu \nab{g}_\mu n_\nu$}  
+\uline{2p    \vp} 
-  n^\nu \nab{\ast}_\nu \rho  
-\uline{p \vp}   -\rho  \vp   
-(p+\rho )  \nab{\Ga}_\nu n^\nu=\cr
 = &
(p-\rho)    \vp
-  n^\nu \nab{\ast}_\nu \rho     
-(p+\rho )  \nab{\Ga}_\nu n^\nu=0   
\cr
}
\fn$$
Now we can eliminate from \ShowLabel{ConservationT} the divergence $\nab{\Ga}_\nu n^\nu$
$$
\eqalign{
\nab{\Ga}_\nu T^{\mu\nu} = &
\nab{\ast}_\nu p g^{\mu\nu} 
-2p  A^\mu + 
\nab{\ast}_\nu (p+\blue{\rho} ) n^\mu n^\nu 
+(\uline{p}+\red{\rho} ) \vp  n^\mu 
+ n^\mu (\uline{p}-\red{\rho})    \vp
- \blue{n^\mu  n^\nu \nab{\ast}_\nu \rho}=\cr
 = &
\nab{\ast}{}^\mu p 
-2p  A^\mu + 
n^\mu n^\nu  \nab{\ast}_\nu p 
+2\vp p  n^\mu =0\cr
&\then
\(g^{\mu\nu}+ n^\mu n^\nu  \) \nab{\ast}_\nu p 
 =2p\( A^\mu -\vp  n^\mu\)
=2p\(  n^\nu \nab{g}_\nu n^\mu  + \red{\vp n^\mu} -\red{\vp  n^\mu}\)=
2p  n^\nu \nab{g}_\nu n^\mu 
\cr
}
\fn$$

Hence conservation of energy-momentum tensor is equivalent to the conditions
$$
\cases{
&(p+\rho )  \nab{\Ga}_\nu n^\nu=(p-\rho)    \vp-  n^\nu \nab{\ast}_\nu \rho  \cr
&\(g^{\mu\nu}+ n^\mu n^\nu  \) \nab{\ast}_\nu p = 2p  n^\nu \nab{g}_\nu n^\mu \cr
}
\fn$$

For dust ($p=0$) one has
$$
\cases{
&  \nab{\Ga}_\nu \(\rho n^\nu\) =-\rho    \vp  \cr
&0 =0 \cr
}
\fn$$
}

By starting from a different conformal gauge fixing $\tilde g= \Phi^2 \cdot g$ one has  different pressure and density, since physical rods and therefore measures depend on the choice of the conformal factor $\Phi$.
In principle one sets
$\tilde p= \Phi^n p$ and $\tilde \rho= \Phi^n \rho$ (with the power $n$ to be determined later in view of conservation of the number of particles) and the energy momentum tensor is
$$
\tilde T_{\mu\nu} := \tilde p \tilde g_{\mu\nu} + (\tilde p+ \tilde \rho ) \tilde n_\mu \tilde n_\nu=
\Phi^{n+2} \( p g_{\mu\nu} + (p+ \rho ) n_\mu n_\nu \)= \Phi^{n+2} T_{\mu\nu}
\fn$$
Of course, unless $n=-2$ (as we shall see below this happens in dimension $m=3$), 
the energy-momentum tensor is not conformally invariant and, more importantly,
its conservation is not preserved by conformal transformations.
If $T_{\mu\nu}$ is conserved then in general $\tilde T_{\mu\nu}$ is not.
Accordingly, one has to specify in which gauge the conservation of energy-momentum tensor has to be imposed.
We shall discuss this issue below in greater detail.

\ss

Also conservation of number of particles can be discussed at kinematical level.
If $\rho$ is related to the density of particles of the fluid, then one would like to define a quantity $J^\mu$ which can be integrated on a spatial region $\Si$ to
determine the number $N_\Si$ of fluid particles hitting $\Si$. The number  $N_\Si$ must be constant along the flow of $n$.

An object to be integrated along an hypersurface $\Si$ is a $(m-1)$-form. There is a natural choice:
$$
J= \sqrt{g} T^{\mu\nu} n_\nu ds_\mu=: J^\mu ds_\mu
\fn$$
and
$$
N_\Si:=\int_\Si   J^\mu ds_\mu 
\fn$$

The number of particles $N_\Si$ is conserved along the flow of $n$ iff one has $d  J =0$.

\CNote{
Let $\Phi_s$ be the flow of $n$ and define a family of hypersurfaces $\Si_s = (\Phi_s)_\ast \Si$.
Then we can define
$$
N_{\Si_s}= \int_{\Si_s}   J^\mu ds_\mu 
= \int_{\Si}  (\Phi_s)^\ast \(J^\mu ds_\mu\)
\fn$$
which allows us to control the variation of $N_{\Si_s}$ along the flow of $n$ by means of the Lie derivative
$$
\eqalign{
\dot {\(N_{\Si_s}\)} |_{s=0}=& \int_{\Si}  \Lie_n \(J^\mu ds_\mu\)
=  \int_{\Si}    d i_n J +  i_n  d J=
\cr
=& \int_{\Si}  \(\del_\la[ u^\la J^\mu]  - \del_\la[ u^\mu J^\la] +   u^\mu \del_\la  J^\la \)ds_\mu=\cr
=& \int_{\Si} [\del_\la u^\la J^\mu +  u^\la \del_\la J^\mu   - \del_\la u^\mu J^\la  - \red{u^\mu \del_\la J^\la}  +   \red{u^\mu \del_\la  J^\la} ]ds_\mu=\cr
=& \int_{\Si} [\del_\la u^\la J^\mu +  u^\la \del_\la J^\mu   - \del_\la u^\mu J^\la ]ds_\mu\cr
}
\fn$$
Now let us notice that 
$$
J^\mu = \sqrt{g} \(p g^{\mu\nu} n_\nu - (p+\rho) n^\mu \)= -\sqrt{g}  \rho n^\mu =: \hat \al n^\mu 
\fn$$
Then one has
$$
\eqalign{
\dot {\(N_{\Si_s}\)} |_{s=0}=& \int_{\Si} [\del_\la n^\la \hat \al n^\mu +  n^\la \del_\la (\hat \al n^\mu)   - \del_\la n^\mu \hat \al n^\la ]ds_\mu=\cr
=& \int_{\Si} [\del_\la n^\la \hat \al n^\mu +  n^\la \del_\la \hat \al n^\mu + \red{ n^\la \hat \al \del_\la n^\mu}
  - \red{\del_\la n^\mu \hat \al n^\la} ]ds_\mu=\cr
=& \int_{\Si} \del_\la (n^\la \hat \al) n^\mu   ds_\mu=\int_{\Si} \del_\la J^\la n^\mu   ds_\mu\cr
}
\fn$$
Hence $N_{\Si_s}$ is constant iff $\del_\mu J^\mu =0$ which is covariant since $J^\mu$ is a vector density of weight $1$.
}

The current $J$ is sensible to conformal transformations since measures are. It can be redefined out of any conformal framework
and one has
$$
\tilde J= \sqrt{\tilde g} \tilde T^{\mu\nu} \tilde n_\nu ds_\mu=  \Phi^m \sqrt{g} \Phi^{n-2} T^{\mu\nu} \Phi n_\nu ds_\mu
= \Phi^{m+n-1} J
\fn$$
The integral $N_\Si$ counts how many fluid particles hit the region $\Si$ and as such Physics imposes that it has to be independent of the conformal framework.
This forces $n$ to be fixed as $n=1-m$ so that $J$ is conformally invariant and its integral is accordingly invariant.

This thence prescribes the following conformal transformations
$$
\cases{
&\tilde p=  \Phi^{1-m} p\cr
&\tilde \rho=  \Phi^{1-m}  \rho\cr
}
\fn$$
This could be expected since $\rho$ and $p$ denotes the {\it spatial densities} (of particles and pressure).

\NewSection{Conservation of $T_{\mu\nu}$}

In a general Weyl context the conservation of energy-momentum tensor and its relation with conservation of the current $J$
needs to be deeply reviewed. 
One has in fact from conservation of number of particles:
$$
\del_\mu J^\mu =\nab{\ast}_\mu J^\mu = \nab{\Ga}_\mu \( \sqrt{g} T^{\mu\nu}n_\nu\)=
 \sqrt{g}  \( \nab{\Ga}_\mu   T^{\mu\nu} n_\nu 
 +   T^{\mu\nu}  \nab{\Ga}_\mu n_\nu \)
 + \nab{\Ga}_\mu \sqrt{ g} \>T^{\mu\nu}n_\nu=0
\fn$$
In standard GR one has $\Ga=\{g\}$ and hence $ \na_\mu \sqrt{ g}=0$; if $n$ is a Killing vector also $T^{\mu\nu}  \na_\mu n_\nu=0$ and conservation of particles, $dJ=0$, is implied by conservation of the energy-momentum stress tensor (which, usually, is eventually implied by Bianchi identities).
However, we have to observe that conservation of $J$ does not on the contrary imply conservation of
energy-momentum tensor. Only the projection of conservation laws along $n$ is in fact used.
Moreover,  being $n$ a Killing vector is a sufficient but by no means necessary condition. 

In a general Weyl context one has in fact
$$
\del_\mu J^\mu =
 \sqrt{g}  \( \nab{\Ga}_\mu   T^{\mu\nu} n_\nu 
 +   T^{\mu\nu}  \nab{\Ga}_\mu n_\nu \)
 +  m\sqrt{ g} A_\mu    \>T^{\mu\nu}n_\nu=0
\fn$$

Thence one should somehow impose that in general one has
$$
T^{\mu\nu}  \nab{\Ga}_\mu n_\nu =0
\qquad
   T^{\mu\nu}A_\mu n_\nu=0
\fn$$
without being too demanding on the vector $n$.

Here the situation is easier for dust ($p=0$). Accordingly, let us first consider this case. 
$$
\eqalign{
T^{\mu\nu}  \nab{\Ga}_\mu n_\nu 
=& \rho n^\mu n^\nu  \nab{\Ga}_\mu \( n^\al g_{\al\nu} \)=\cr 
=& \rho \(n_\al  n^\mu  \nab{\Ga}_\mu n^\al  - 2  n^\mu A_\mu \)
= \rho(-\vp   +2  \vp)=\vp\rho
}
\fn$$
Now we have to stress that the scalar $\vp$ can be chosen at will. Here of course we would like to fix $\vp=0$. This corresponds to require that the integral curves of $n$ are not only geodesics trajectories of $\Ga$,
but in fact they are geodesics motions. We here started to be general enough but we are forced back to geodesic motions.

As far as the second condition is concerned one has
$$
    T^{\mu\nu}A_\mu n_\nu = -\rho A_\mu  n^\mu  =
  \rho    \vp    =0
\fn$$
which is also satisfied for $\vp=0$.

The situation with pressure is more complicated.
$$
\eqalign{
T^{\mu\nu}  \nab{\Ga}_\mu n_\nu =& pg^{\mu\nu}\nab{\Ga}_\mu n_\nu + (p+\rho )n^\nu  n^\mu  \nab{\Ga}_\mu n_\nu =\cr
=& p\nab{\Ga}_\mu n^\mu  + 2p n^\mu A_\mu
  + (p+\rho )n_\al  n^\mu  \nab{\Ga}_\mu n^\al
  -  2(p+\rho )  n^\mu  A_\mu=\cr
=& p\nab{\Ga}_\mu n^\mu  - 2p \vp
  - (p+\rho ) \vp 
  + 2(p+\rho )  \vp= p\nab{\Ga}_\mu n^\mu  + 2\rho  \vp =\cr
=&  p\nab{g}_\mu n^\mu + pn^\la\(g^{\mu\ep} g_{\la\mu}- \de^\mu_{\la}\de^\ep_{\mu}- \de^\mu_{\mu}\de^\ep_{\la}\) A_\ep + 2\rho  \vp =\cr
=&  p\nab{g}_\mu n^\mu + p n^\la\(\red{\de^\ep_\la} - \red{\de^\ep_{\la}}- m\de^\ep_{\la}\) A_\ep + 2\rho  \vp =
p\nab{g}_\mu n^\mu  +\( 2\rho-mp  \) \vp \cr
}
\fn$$
and one has to face the incompressibility condition $\nab{g}_\mu n^\mu =0$.
This can be solved by using the freedom in the conformal gauge fixing.
One can show that there is always a conformal representative for which
$$
\nab{\tilde g}_\mu \tilde n^\mu=0
\fn$$
and then in this conformal frame we fix $\tilde \vp=0$.

\Note
One has
$$
\eqalign{
\nab{\tilde g}_\mu \tilde n^\mu=&  \nab{g}_\mu \tilde n^\mu -  \tilde n^\la \( g^{\mu\ep} g_{\la\mu} - \de^\mu_\la \de^\ep_{\mu} - \de^\mu_\mu \de^\ep_{\la}\) \del_\ep\ln \Phi=\cr
=&  \nab{g}_\mu \tilde n^\mu - \tilde n^\la\( \red{\de^\ep_\la}  - \red{\de^\ep_\la}  - m  \de^\ep_{\la}\) \del_\ep\ln\Phi =
\nab{g}_\mu \tilde n^\mu + m    \tilde n^\ep \del_\ep\ln \Phi=\cr
=&  \frac[1/\Phi]  \(\nab{g}_\mu n^\mu + (m-1)    n^\ep \del_\ep\ln \Phi\) \cr
}
\fn$$

Now, the fact is that whaterver $\nab{g}_\mu n^\mu$ one can always find a conformal factor $\Phi$ such that
$$
 n^\ep \del_\ep\ln \Phi= - \Frac[1/m-1]\nab{g}_\mu n^\mu 
\fn$$
(fix coordinates in which $n= \del_0$).
Using such a conformal factor to change conformal representative, in the new conformal frame one has
$\nab{\tilde g}_\mu \tilde n^\mu=0$.
\endNote

For the second condition to hold one has
$$
 \eqalign{
  T^{\mu\nu}A_\mu n_\nu=&
-(\rho+ p)  A_\mu n^\mu  + pA_\mu n^\mu=
-\rho  A_\mu n^\mu  =
\rho   \vp   
 }
\fn$$
which also vanishes under the same condition.

Thus in general one needs not to require that $n$ is Killing. 
In fact the fluid (with pressure) selects a preferred conformal frame $\tilde g$.
In that preferred frame  one has a preferred Weyl connection with 
$$
\tilde A_\nu = \tilde n^\mu \nab{\tilde g}_\mu \tilde n_\nu 
\fn$$
for which $\tilde n^\mu \nab{\Ga}_\mu \tilde n^\nu =0$.
With these choices, not only the fluid velocities are a geodesic field, but the conservation law $d J=0$ is equivalent to the conservation of fluid energy-momentun tensor, $\nab{\Ga}_\mu T^{\mu\nu}=0$.

\NewSection{Conclusions and Perspectives}

We showed that for any conformal structure $[g]$ on a spacetime $M$ and for any timelike vector field $u$
one can always determine an EPS-compatible connection $\Ga$ for which the vector field $u$ is geodesic.
Then one can determine the conformal frame and parametrizations along worldlines so that
one has the standard relation among the different conservations associated to the fluid.

Of course we are not here suggesting that the model we presented here is physically sound. 
One should specify a dynamics and then investigate the relation with Bianchi identities.
We chose this model to show that had Einstein known it, he could have possibly  tried this way to model matter and gravity and make them  compatible with SR.

This research is part of a larger project aiming to model a generic self-gravitating fluid in EPS formalism.
One can specify the conformal structure to be the Minkowskian one by setting $g=\eta$ and still be free to
model any congruence of (timelike) worldlines as the flow of a fluid. In this framework the gravitational field is encoded into the covector $A$ which in turn determines the Weyl connection $\Ga$.

Let us stress that in the EPS setting there is no freedom in choosing the connection associated to free fall. In EPS framework the free fall of particles is {\it by construction} described by the connection $\Ga$ while
the Levi-Civita connection of $g$ plays just the kinematical role of {\it reference frame} in the affine space of connections (which is moreover conformally covariant since one can start from any representative of the conformal structure).

The extra degrees of freedom to determine $\Ga$ are thence encoded into the covector $A$ which is kinematically free to be generic.
The dynamics of the theory determines then the connection $\Ga$ fixing the covector $A$ in terms of matter fields and $g$.

Of course Weyl geometries are affected by physical interpretation problems mainly related to the (possibly non-trivial) holonomy of the connection $\Ga$; see \ref{EPS}. 
However, these problems arise only if $\Ga$ is not metric while metric connections do not generate any physical problem of this kind and have just to be interpreted correctly. We stress that in all $f(R)$ models with non-exotic matter (or in vacuum) dynamics forces a posteriori the connection $\Ga$ to be automatically metric (and in fact to be the Levi-Civita connection of a metric conformal to the $g$ originally entering the Lagrangian; see \ref{EPS1} and \ref{EPS2}),
unless one introduces a matter Lagrangian in which matter couples directly with the connection
$\Ga$ (a case in which the connection can be also not metric; see \ref{MGaCou}).

\Acknowledgements
This work is supported by INFN (Iniziativa Specifica NA12) and by INdAM-GNFM.
We wish to thank N.Fornengo for discussions about geodesics and fluids.
We also wish to thank M.Ferraris, G.Magnano and P. Menotti for discussions and comments.

\ShowBiblio

\end